\documentclass[twocolumn,showpacs,preprintnumbers,amsmath,amssymb]{revtex4}

\usepackage{dcolumn}
\usepackage{bm}
\usepackage{epsfig}

\begin{document}


\title{Evidence for nonmonotonic magnetic field penetration in a type-I superconductor}

\author{V.F. Kozhevnikov$^{1*}$, C.V. Giuraniuc$^1$, M.J. Van Bael$^1$, K. Temst$^2$,  C. Van Haesendonck$^1$,
T.M. Mishonov$^3$, T. Charlton$^4$, R.M. Dalgliesh$^4$, Yu.N.
Khaidukov$^5$, Yu.V. Nikitenko$^5$, V.L. Aksenov$^5$, V.N.
Gladilin$^{1,6}$, V.M. Fomin$^{6}$, J.T. Devreese$^{6}$,  and J.O.
Indekeu$^1$ }

\affiliation{
$^1$Laboratorium voor Vaste-Stoffysica en Magnetisme Katholieke Universiteit Leuven, 3001 Leuven, Belgium \\
$^2$Instituut voor Kern- en Stralingsfysica, Katholieke Universiteit Leuven, 3001 Leuven, Belgium \\
$^3$Department of Theoretical Physics, St Clement of Ohrid University of Sofia, 1164 Sofia, Bulgaria\\
$^4$ISIS Science Division, Rutherford Appleton Laboratory, Chilton, Didcot OX11 0QX, United Kingdom \\
$^5$Frank Laboratory of Neutron Physics, Joint Institute for Nuclear Research, 141980 Dubna, Moscow Region, Russia \\
$^6$Theoretische Fysica van de Vaste Stoffen, Universiteit Antwerpen, 2020 Antwerpen, Belgium}. \\

\date{\today}

\begin{abstract}
\noindent Polarized neutron reflectometry (PNR) provides evidence
that {\em nonlocal} electrodynamics governs the magnetic field
penetration in an extreme low-$\kappa $ superconductor. The sample
is an indium film with a large elastic mean free path (11 $\mu$m)
deposited on a silicon oxide wafer. It is shown that PNR can
resolve the difference between the reflected neutron spin
asymmetries predicted by the local and nonlocal theories of
superconductivity. The experimental data support the nonlocal
theory, which predicts a {\em nonmonotonic decay} of the magnetic
field.
\end{abstract}


\maketitle

In this paper we pose and answer experimentally the following
fundamental questions. Are nonlocal electrodynamics effects
measurable in superconductors? Can the nonmonotonic decay of
magnetic field penetration predicted by the nonlocal theory be
observed? To what extent can Polarized Neutron Reflectometry (PNR)
resolve the difference between local and nonlocal diamagnetic
responses expected for strongly type-I superconductors?

{\em Nonlocality} is a key concept of superconductivity theory,
but its experimental verification is still not established. In the
Meissner state, a magnetic field applied parallel to the surface
located at $z=0$ causes the magnetic induction $B(z)$ to penetrate
over a depth $\lambda \equiv B(0)^{-1}\int B(z) dz$. In the London
({\em local}) limit, appropriate to most type-II superconductors,
$B(z)\propto \exp(-z/\lambda_L)$, where $\lambda_L$ is the London
penetration depth. In 1953, to explain the variation of $\lambda$
in type-I superconductor Sn caused by adding In, Pippard proposed
that the current density is related to the average of the vector
potential over a region of size $\xi_0$ (the Pippard coherence
length) \cite{Pippard}. The smaller the Ginzburg-Landau parameter
$\kappa$, the more important this {\em nonlocal} effect.

Nonlocal theory predicts that $B(z)$ deviates from a simple
exponential decay. $B(z)$ is {\em nonmonotonic} and, moreover,
{\em changes sign} at a specific depth \cite{Pippard}. In the pure
limit ($\xi_0 \ll \ell$, where $\ell$ is the elastic mean free
path) $B(z)$ is determined by the intrinsic parameters
$\lambda_L(T=0)$ and $\xi_0$, and by the temperature $T$. The
deviation is most significant in ``extreme" type-I superconductors
($\kappa \ll 1/\sqrt{2}$), such as Al ($\kappa\approx 0.01$) and
In (0.06). For these the results of the Pippard theory are
identical to those of the Bardeen-Cooper-Schrieffer theory
\cite{Tinkham}. On the other hand, the local approximation is safe
for $\kappa > 1.5$ \cite{Halbritter}, i.e., marginal nonlocal
effects are predicted for some type-II superconductors such as Nb
($\kappa \approx 1$).

$B(z)$ in In, calculated in local and nonlocal approaches, in the
pure limit with $\xi_0$ = 0.38 $\mu$m and $\lambda_L(0)$ = 0.025
$\mu$m \cite{Valko}, for $T$ = 1.8 K, is shown in Fig.\,1. Details
about the formalism can be found in Ref.\,\onlinecite{Halbritter}.
In the nonlocal approach $B(2\lambda_L)/B(0)\approx 1/e$. The sign
reversal is expected at $z \approx $ 5.5$\, \lambda_L$, and the
amplitude of the reversed field is at most about 3$\%$ of the
field at the surface.
\begin{figure}
\epsfig{file=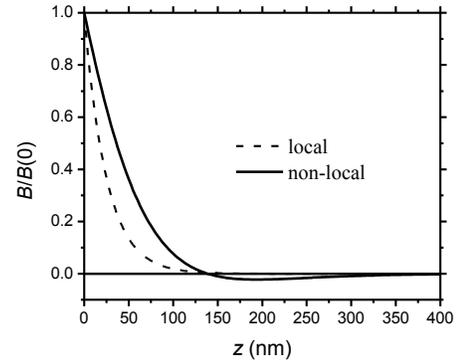,width=6 cm}
\caption{\label{fig:epsart} Magnetic-induction profiles $B(z)$ in
a semi-infinite In sample. The dashed (solid) line corresponds to
the local (nonlocal) relation between current density and vector
potential.}
\end{figure}

An observation of sign reversal was reported in Ref.\,
\onlinecite{Drangeid}. An external AC magnetic field $H$ with
amplitude up to 30 Oe was applied parallel to a hollow cylindrical
Sn film about 2 $\mu$m thick, and a strongly attenuated ($10^8$
times) signal with reversed phase was detected inside the cylinder
at $T$ = 2.88 K and $H \approx$ 25 Oe. This phase difference was
interpreted as a change of sign in the penetrating field. However,
this is questionable because the phase difference drops back to
zero at a larger (30 Oe) field, whereas the critical field $H_c$
at 2.9 K is 115 Oe \cite{Finnemore_In}.

Nowadays $B(z)$ can be measured directly using polarized neutron
reflectometry (PNR) \cite{Daillant} and low-energy muon spin
rotation (LE-$\mu$SR) \cite{Morenzoni} techniques. We comment
briefly on the latter before focusing on the former.

In the LE-$\mu$SR technique positive muons polarized
perpendicularly to the applied field are implanted in a sample
over a distance determined by the muon energy. $B(z)$ is obtained
from measuring the Larmor precession frequency of the muon spins
at stopping distance. In other words, the implanted muons serve as
tiny sensors of the magnetic field inside the sample. However, in
practice the muon precession is progressively damped with depth
due to a broad distribution of stopping distances \cite{Suter}.
From our prospectus, this is the main difficulty in applying the
LE-$\mu$SR technique to fields with a sharp profile.

Recently, the LE-$\mu$SR technique was used to measure $B(z)$ in
Pb, Nb, and Ta \cite{Suter}. Most interesting is the observation
of a non-exponential shape of $B(z)$ for all studied metals. The
nonlinearity of the semi-log plots for $B(z)$ is marginal, which
is exactly what should be expected theoretically in view of the
fairly high $\kappa$ of the studied samples. For example, $\kappa$
of pure Nb (residual resistivity ratio $RRR = 1600$) is 1.3 at 3 K
and 1.0 at 7 K \cite{Finnemore}. However, in
Ref.\,\onlinecite{Suter} $\kappa$ of less pure Nb ($RRR$=133) is
reported to be 0.7 at 2.96 K and 7.6 K. This and some other
inconsistencies with well established literature data suggest that
the muon probing results may contain some hidden uncertainties.
Therefore, additional experiments would be worthwhile, in
particular on low-$\kappa$ superconductors.

The PNR technique is based on the change of the neutron index of
refraction in a magnetized medium. When a collimated neutron beam
polarized along the magnetic field is incident on a flat,
laterally uniform sample under a grazing angle, its specular
reflectivity $R$ is determined by the profile of the neutron
scattering potential below the surface. $R$ is measured versus
momentum transfer $Q=4\pi\sin\theta/\lambda_n$, where $\theta$ is
the angle of incidence and $\lambda_n$ the neutron wavelength. The
scattering potential consists of a nuclear and a magnetic part,
which results in different reflectivities $R^+$ and $R^-$ for
neutrons with spins polarized parallel (up) and anti-parallel
(down) to the applied field, respectively. Direct information
about the sample magnetization is obtained by combining $R^+$ and
$R^-$; the combination $s=(R^+-R^-)/(R^++R^-)$ is the spin
asymmetry. $B(z)$ can be found by fitting $s(Q)$ data with $s(Q)$
calculations based on theoretical models for $B(z)$. PNR has been
applied for measuring the penetration depth and for detecting
surface superconductivity in Nb \cite{Felcher, Zhang}, high-T$_c$
cuprates \cite{Felici, Masour, Lauter-Pasyuk} and Pb \cite{Gray,
Nutley}.

The nonlocal effect in $B(z)$ measured with PNR was discussed in
Refs.\,\onlinecite{Felcher}, \onlinecite{Zhang}, \onlinecite{Gray}
and \onlinecite{Nutley}. Although some deviation from exponential
decay was noticed in Refs.\,\onlinecite{Zhang} and
\onlinecite{Gray}, no solid confirmation of the nonlocal theory
was obtained. The authors of Ref.\,\onlinecite{Nutley} correctly
pointed out that experiments with low-$\kappa$ type-I
superconductors are desirable to verify nonlocality, but their
overall conclusion was that PNR is incapable of detecting
nonlocality in {\em any} superconductor. In this light it is very
interesting to reassess the problem of nonlocality with
state-of-the-art PNR applied to a low-$\kappa$ material such as
In, since during the last decade many neutron source facilities
have significantly progressed in neutron flux, neutron optics and
detector technology.

The design of the sample for the PNR study is based on the
following requirements. The irradiated surface must be flat and
possess minimal possible roughness. The sample must be thick
enough to have the same electromagnetic properties as the bulk
material. Degradation of the surface quality with increasing
thickness limits the film thickness. Neutrons reflected back from
the substrate should have a negligible effect on the reflectivity
in a region close to the critical edge of total reflection, $Q_c$,
where the reflectivity is most sensitive to the magnetic
properties.

Two approaches can meet these requirements. One is to deposit a
thick film on a flat substrate that reflects least. This can be
achieved if the neutron refraction index of the substrate is
larger than that of the sample. This approach was taken in the
experiments on Nb \cite{Felcher, Zhang} and Pb \cite{Gray,
Nutley}. In fact, this was the only option, in view of the
negligibly small absorption of neutrons in Nb and Pb. However, In
is a strong absorber, which enables one to rely on substrates with
a refractive index smaller than that of In, provided the thickness
of the indium film is properly optimized. In this approach a
second plateau or ``hill", associated with total reflection from
the sample-substrate interface, is expected in the reflectivity
curve $R(Q)$. This should yield additional information about the
sample structure. Modelling shows that an indium thickness of 2.5
$\mu$m is appropriate. Such a sample was fabricated in the present
work.

High purity indium (99.9999$\%$) was deposited by thermal
evaporation on the polished side of a silicon oxide wafer at room
temperature. The substrate size was 2$\times$2 cm$^2$$\times$1 mm.
The base pressure and the evaporation rate were 4$\times$10$^{-8}$
mbar and 60-70 \AA/s, respectively. The
nominal film thickness, as recorded by a
quartz
monitor, was 2.5 $\mu$m. Several smaller area samples were
simultaneously fabricated for the film characterization.

The root-mean-square (rms) surface roughness $\sigma$ probed with
an atomic force microscope (AFM) yielded 2.0, 6.7 and 8.0 nm at
the scale of 1, 5 and 10 $\mu$m, respectively. A scan range up to
10 $\mu$m was not sufficient to reach saturation of the roughness.
Consequently, 8.0 nm is a lower bound on the roughness at the
scale of the neutron coherence length ($\approx$ 100 $\mu$m
\cite{Temst}). In our simulations, effects due to surface
roughness are modelled using N\'{e}vot-Croce factors
\cite{Daillant}, where the roughness is characterized by $\sigma$;
it was allowed to vary to fit the experimental data.

Another parameter associated with the sample surface is the
thickness of the indium oxide film. When exposed to air, In, like
its neighbors in the Periodic Table, Al and Ga, instantly forms a
protective oxide layer. A surface of indium in air remains
lustrous for years. This suggests that the oxide layer is very
thin, perhaps of the order of a few monolayers, and should not
affect the neutron reflectivity. This is consistent with the
negative result of Rutherford backscattering measurements
performed on our sample: no oxide film has been detected.

The electromagnetic properties of the sample were characterized by
measurements of the DC magnetization $M$ and the electrical
resistivity. The shape of the $M(H)$-curves is typical for type-I
superconductors. The obtained phase diagram $H_c(T)$ agrees well
with the literature data \cite{Finnemore_In}. $T_c$ of our sample
(3.415 K) matches the tabulated value of 3.4145 K
\cite{Grigoriev}, and $RRR = 540$. Correspondingly, $\ell \approx
11 \, \mu {\rm m}$ is much larger than $\xi_0$. Therefore, our
sample is a type-I superconductor in the pure limit.

PNR experiments were performed both on the REMUR reflectometer
\cite{REMUR} at the Joint Institute for Nuclear Research (Dubna)
and on the CRISP instrument \cite{CRISP} at ISIS (Oxford). Both
sets of measurements confirm that splitting of the $R^+(Q)$ and
$R^-(Q)$ curves is achievable for our sample. The ISIS data, which
are the most detailed, allow a quantitative analysis to which we
now turn.

CRISP operates with a spin-polarized polychromatic pulsed neutron
beam. The angle of incidence and the instrumental resolution
$\Delta Q/Q$ were set to 0.24 degrees and 3$\%$, respectively.

The reflectivity in the Meissner state was measured at $T$ = 1.8 K
and at magnetic fields of 77, 140, 166, and 194 Oe ($H_c$(1.8 K) =
205 Oe). The obtained data sets are shown in Fig.\,2. The
$R(Q)$-dependencies exhibit a hill caused by the total reflection
from the substrate. The splitting between $R^+$ and $R^-$ is
clearly visible near $Q_c$; different magnitudes of the error bars
are due to different times of exposure. The data obtained at 77
and 166 Oe have the smallest statistical error and will be used
for further discussion.

\begin{figure}
\includegraphics[width=6.5 cm]{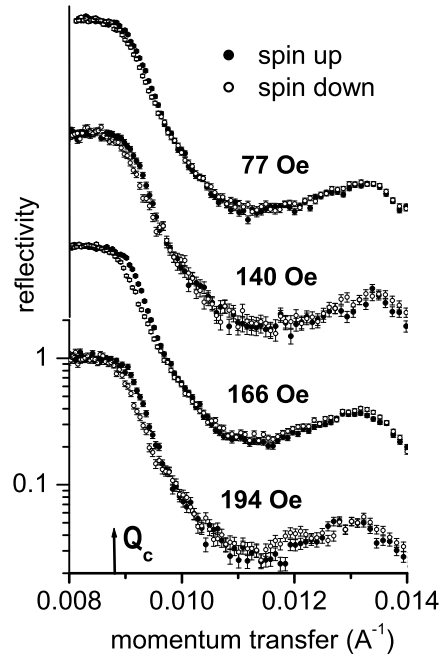}
\caption{\label{fig:epsart} Reflectivity of polarized neutrons
measured in the Meissner state. $Q_c$ is the momentum transfer for
total neutron reflection from the outer surface. The scale is
shown for the data at $H=$ 194 Oe; the other data have been
shifted for clarity.}
\end{figure}

The data for the reflectivity in the normal state are shown in
Fig.\,3. Solid curves are simulations, in which the sample was
represented by a pure In film on a SiO$_2$ substrate. In the
simulations the angular beam resolution was allowed to vary due to
the unknown uncertainties of the instrumental resolution and of
the geometrical factor (as only part of the beam covers the
sample).

The simulation curve near $Q_c$ is mostly controlled by the
resolution (see also \cite{Gray}). The next segment, down to the
foothill, is determined by the roughness of the sample surface.
The location of the ascending part ($0.011 < Q($\AA$^{-1}) <
0.014$) is governed by the film thickness. The curve segment
following the hill is determined by the substrate scattering
properties. No attempts were made to achieve a better fit for that
segment, because there the spin asymmetry is indistinguishable
from zero.

\begin{figure}
\includegraphics[width=6.cm]{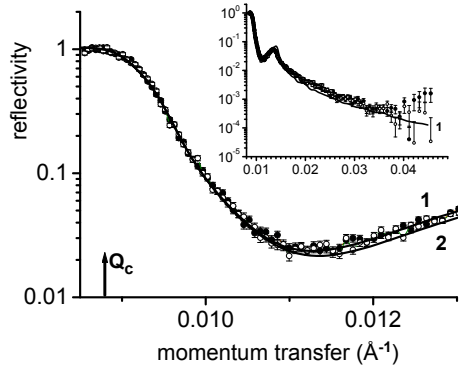}
\caption{\label{fig:epsart} Neutron reflectivity at $T=$ 4.6 K.
Curves 1 and 2 are simulations for a film thickness of 2.40 and
2.50 $\mu$m, respectively. The inset shows the data for the full
range of $Q$-values.}
\end{figure}

The best fit (Fig.\,3) was obtained for the model sample with
$\sigma$ = 14 nm and $\Delta Q/Q$ = 2.5 $\%$. Fitting the
ascending part enables one to determine the film thickness {\em in
situ}. The statistical error of the reflectivity data in this
region being $\pm$ 5$\%$, the thickness was found to be 2400 $\pm$
30 nm, in agreement with the nominal thickness of 2.5 $\mu$m.
These parameters were further used for simulating the spin
asymmetry. Attempts to introduce an indium oxide layer on top of
the sample yielded no reasonable fit for any appreciable thickness
($> 1 $nm) of the oxide layer. This is consistent with our
expectation that the indium oxide layer does not affect the
neutron reflectivity.

\begin{figure}
\includegraphics[width=6.cm]{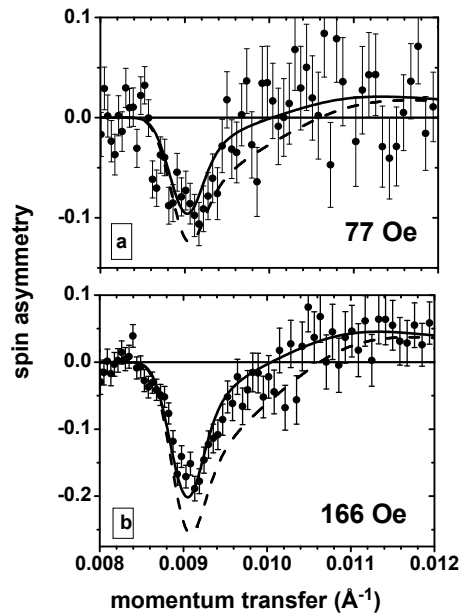}
\caption{\label{fig:epsart} Spin asymmetry at $T=$ 1.8 K and $H=$
77 Oe (a) and 166 Oe (b). The curves are simulations performed
within the local (dashed line) and nonlocal (solid line)
approaches.}
\end{figure}

For our simulations of the reflectivity in the Meissner state, the
magnetic field profiles shown in Fig.\,1 were assumed on both
sides of the sample. The spin asymmetry data for fields 77 Oe and
166 Oe, along with simulations for the local and nonlocal field
distributions, are shown in Fig.\,4.

For field 77 Oe (Fig.\,4a), the results of the ``nonlocal"
simulation fit the experimental data somewhat better, but no clear
discrimination between the local and nonlocal approaches is
possible due to insufficient accuracy of the data at this field. A
significantly clearer distinction is apparent for field 166 Oe due
to the larger amplitude of $s(Q)$. As can be seen from Fig.\,4b,
the quality of the fits, with $B(z)$ calculated in the local and
nonlocal approaches, is different.  The nonlocal simulation fits
the experimental data definitely better. It is worth stressing
that {\em no adjustable parameters} have been used for the
simulations of spin asymmetry.

In conclusion, nonlocal electrodynamics effects are measurable in
extreme type-I superconductors. State-of-the-art PNR measurements
performed on low-$\kappa$ superconductor In, combined with
simulation, unambiguously support the nonlocal theory and at the
same time demonstrate consistency with the literature data for
$\lambda_L(0)$ and $\xi_0$. Consequently, evidence has been
gathered for the nonmonotonic decay and sign reversal of the
penetrating magnetic field predicted by the nonlocal
electrodynamics approach.

We thank A. Volodin for AFM, S. Vandezande for electrical
conductivity, and A.P. Kobzev for Rutherford backscattering
measurements. This research has been supported by the KULeuven
Research Council (F/05/049, GOA/2004/02), project G.0237.05 of
FWO-Vlaanderen, IUAP P5/1, the European Commission 6th Framework
Programme through Key Action: Strengthening the European Research
Area, Research Infrastructures (Contract HII3-CT-2003-505925),
Russian State contract 2007-3-1.3-07-01, INTAS grant 03-51-6426
and RFBR project 06-02-16221.

\bibliography{apssamp}

\end{document}